\documentclass[manuscript]{aastex}

\usepackage{tikz}
\usepackage{amsmath}

\usepackage{graphicx}

\usepackage{pdflscape}

\slugcomment{\textbf{The Astronomical Journal}}

\shorttitle{Long-Period Comet C/2014 B1}
\shortauthors{Jewitt}

\begin{document}

\title{The Discus Comet: C/2014 B1 (Schwartz)}

% Use \author, \affil, and the \and command to format
% author and affiliation information.
% Note that \email has replaced the old \authoremail command
% from AASTeX v4.0. You can use \email to mark an email address
% anywhere in the paper, not just in the front matter.
% As in the title, use \\ to force line breaks.

\author{David Jewitt$^{1,2}$
Yoonyoung Kim$^3$, 
Jane Luu$^{4}$
and
Ariel Graykowski$^1$
} 
\affil{$^1$Department of Earth, Planetary and Space Sciences,
UCLA, 
595 Charles Young Drive East, 
Los Angeles, CA 90095-1567\\
$^2$Department~of Physics and Astronomy,
University of California at Los Angeles, \\
430 Portola Plaza, Box 951547,
Los Angeles, CA 90095-1547\\
$^3$ Max Planck Institute for Solar System Research, Justus-von-Liebig-Weg 3, 37077 G\"ottingen, Germany\\
$^4$ Department of Physics and Technology, Arctic University of Tromso, Tromso, Norway \\
}

\email{jewitt@ucla.edu}

\begin{abstract}
Long period comet C/2014 B1 (Schwartz) exhibits a remarkable optical appearance, like that of a discus or bi-convex lens viewed edgewise.  Our measurements in the four years since discovery reveal a unique elongated dust coma whose orientation is stable with respect to the projected anti-solar and orbital directions.  With no tail and no trail, the limited influence of radiation pressure on the dust coma sets a lower limit to the effective particle size $\gtrsim$100 $\mu$m, while the photometry reveals a peak coma scattering cross-section $2.7\times 10^{4}$ km$^2$ (geometric albedo 0.1 assumed).  From the rate of brightening of the comet we infer a dust production rate $\gtrsim$10 kg s$^{-1}$ at 10 AU heliocentric distance, presumably due to the sublimation of supervolatile ices, and perhaps triggered by the crystallization of amorphous water ice.   We consider several models for the origin of the peculiar morphology. The disk-like shape is best explained by equatorial ejection of particles from a nucleus whose spin vector lies near the plane of the sky.  In this interpretation, the unique appearance of C/2014 B1 is a result of a near equality between the rotation-assisted nucleus escape speed ($\sim$1 to 10 m s$^{-1}$ for a 2 to 20 kilometer-scale nucleus) and the  particle ejection velocity, combined with a near-equatorial viewing perspective.    To date, most other comets have been studied at heliocentric distances less than half that of C/2014 B1, where their nucleus temperatures, gas fluxes and dust ejection speeds are much higher. The throttling role of nucleus gravity is correspondingly diminished, so that  the disk morphology has not before been observed. 
\end{abstract}


\section{INTRODUCTION}

Long period comet C/2014 B1 (Schwartz) (hereafter ``B1'')  follows a hyperbolic orbit  with semimajor axis $a$ = -2160 AU, eccentricity $e$ = 1.00442 and inclination $i$ = 28.4\degr~(orbital elements are from JPL's HORIZONS on-line database).  Although B1 is technically not gravitationally bound to the Sun, the eccentricity excess above $e$ = 1 is so small that an interstellar origin is unlikely.  Instead, along with numerous other  long-period comets having slightly hyperbolic orbits, B1 is  of probable Oort cloud origin.  Perihelion occurred at $q$ = 9.557 AU on UT 2017 September 10.

Apart from its atypically large perihelion distance, the special feature of B1 is its peculiar, indeed unique, optical morphology.  Specifically, its shape resembles  a discus or bi-convex lens viewed edgewise.  Moreover, the  major axis of symmetry of this elongated object lies nearly (but not exactly) perpendicular  to the projected  orbit and shows no relation to the  anti-solar direction.    In other comets, solar gravity and solar radiation pressure are  the dominant forces shaping the large-scale morphology of the dust coma.   Small dust particles are accelerated by the momentum of solar photons and pushed into a radiation pressure swept tail that is aligned approximately in the anti-solar direction.  Large particles are less influenced by radiation pressure and are also ejected more slowly from the nucleus; they instead populate a narrow region along the orbital plane, appearing as a band or ``trail'' in the plane of the sky.  The morphology of B1, and the lack of a clear relation to the anti-solar and projected orbit directions together present a puzzle regarding the nature of the dust and the mechanism of its ejection.

In this paper we present observations taken to characterize B1 and to attempt to understand its unique and distinctive appearance.

\section{OBSERVATIONS}
Observations were acquired using the Wisconsin-Indiana-Yale-NOAO (WIYN) 0.9 m telescope, the Nordic Optical Telescope (NOT) 2.5 m,  %the WIYN 3.5 m telescope 
and the Keck 10 m  telescope.  A journal of observations, including basic instrumental parameters and the observing geometry for each date of observation, is given in Table (\ref{geometry}).   In the Table, we express dates as DOY (Day of Year) where DOY = 1 on UT 2014 January 1.  For reference,  B1 was discovered on DOY = 28 (Schwartz and Sato 2014) and perihelion occurred on DOY = 1349. 

\begin{figure}[ht]
\centering
\includegraphics[width=0.95\textwidth]{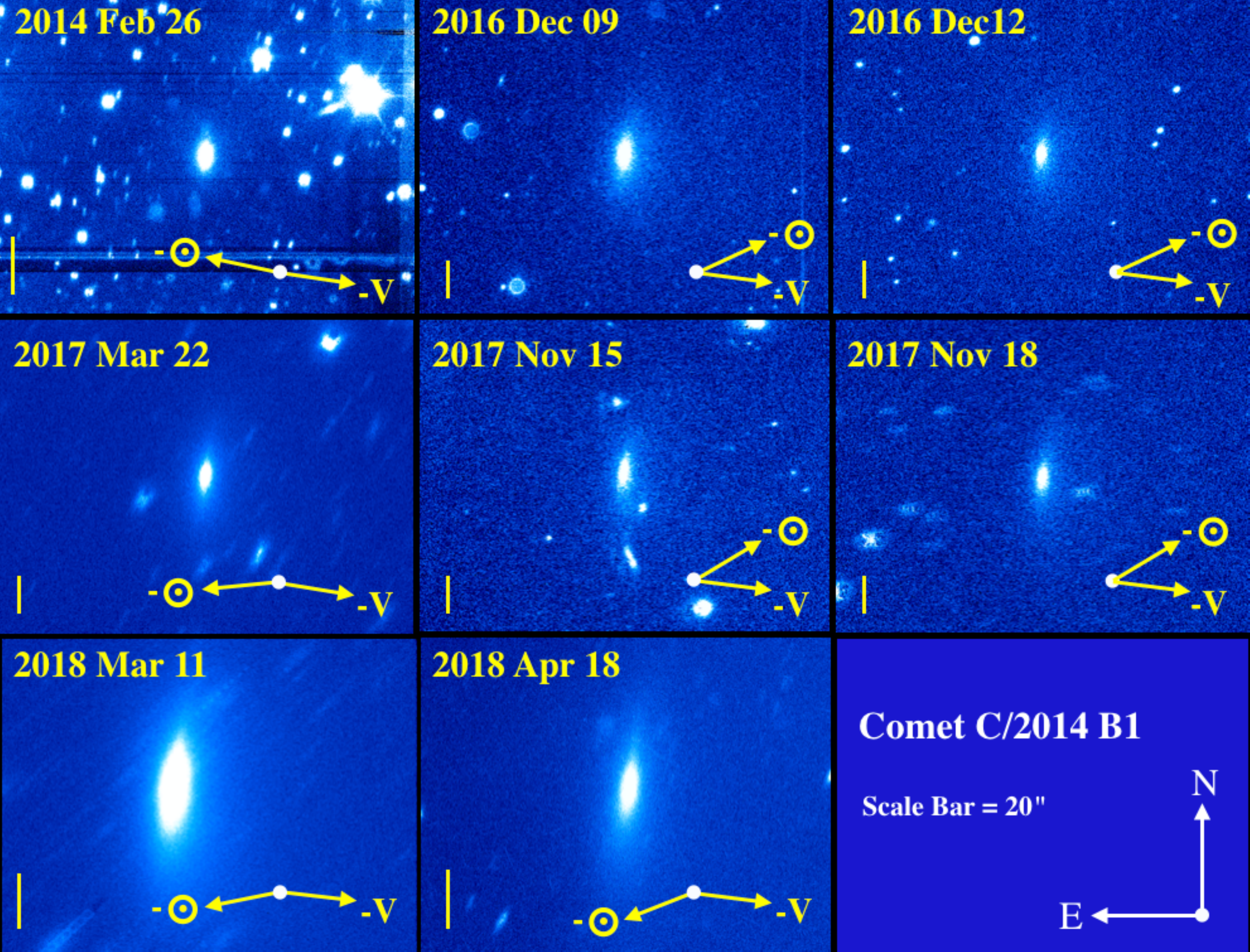}
\caption{Image composites corresponding to Table (\ref{geometry}) arranged in order of date and showing the fixed orientation of the major axis of B1.  In each panel we show the directions of the projected antisolar vector ($-{\odot}$) and the negative projected heliocentric velocity vector ($-V$) in yellow.  The vertical bar denotes 20\arcsec.  The cardinal directions are marked in the lower right. \label{composite} }
\end{figure}

We corrected for spatial sensitivity variations across the charge-coupled detectors employed at each telescope, using  flat fields computed from the median of a large number of uncorrelated night sky images (at Keck) or obtained from an illuminated patch on the inside of the telescope dome (at the other telescopes).  Photometric calibration of the red filter data was obtained from observations of Sun-like stars from the catalog by Landolt (1992) and from field objects measured in the Sloan Digital Sky Survey Data Release 14 (Abolfathi et al.~2018).  Magnitudes obtained using Sloan filters were transformed to the Johnson-Cousins BVR system using the relations determined by Jester et al.~(2005).  In both cases, the standard star data are accurate to $\sim \pm$0.01 magnitudes in BVR and the photometric stability of the atmosphere was typically $\pm$0.01 to $\pm$0.02 magnitudes. The accuracy of the comet photometry (Table \ref{photometry}) is further limited by uncertainty in the sky background owing to the presence of field stars and galaxies.  We present a composite of the data from each night listed in Table (\ref{geometry}) in Figure (\ref{composite}).  These images are from telescopes with widely different collecting areas, have different total integration times and resulting sensitivities, and were taken in different sky conditions. While the extent of the measurable coma varies from image to image, the overall stability of the morphology is noteworthy, particularly with respect to the varying projected anti-solar vector, marked in yellow in each panel.  A high quality  image from the NOT telescope is shown in Figure (\ref{image}).

\subsection{Photometry}
We took measurements using an aperture with an angular radius scaled inversely with geocentric distance ($\Delta$, see Table \ref{geometry}), so as to always measure the light scattered from within a fixed linear (not angular) distance from the nucleus (which we assumed to be located at the opto-center of the comet).  In this way, the measured apparent brightness of a fixed cross-section coma should vary in accordance with the inverse square law without need for an additional geometric correction related to the surface brightness distribution in the coma.  Table (\ref{photometry}) lists apparent magnitudes obtained within a circular aperture of projected radius 4$\times$10$^4$ km at the comet, denoted $m_R(40)$.   If there were no change in the coma of B1 then $m_R(40)$ would vary according to the inverse-square law, here expressed as 

\begin{equation}
m_R(40) = H_R(40) + 2.5\log_{10}\left(r_H^2 \Delta^2\right) -  2.5\log_{10}(\Phi(\alpha))
\label{abs}
\end{equation}

\noindent where $r_H$ and $\Delta$ are the heliocentric and geocentric distances expressed in AU and $\Phi(\alpha) \le 1$ is the phase function at phase angle $\alpha$.  Equation (\ref{abs}) defines the  absolute red magnitude,  $H_R(40)$, as the magnitude the comet would have if it could be observed from $r_H = \Delta = 1$ AU and $\alpha =$ 0\degr.  The phase functions of comets are reasonably well-approximated by simple functions of $\alpha$, provided $\alpha \lesssim$ 50\degr.   We write $-2.5\log_{10}(\Phi(\alpha))$ =  $k \alpha$, where $k$ is a constant.   The phase functions of cometary nuclei fall in the range 0.02 $\le k \le$ 0.1 (Kokotanekova et al.~2017).  The phase functions measured in active near-Sun comets tend to be smaller; we take $k$ = 0.4 magnitudes degree$^{-1}$ (Meech and Jewitt 1987).  Over the modest range of phase angles taken by B1 ($\alpha \le$ 3.8\degr, Table \ref{geometry}), even a $\pm$50\% uncertainty in the coefficient would have only a modest effect ($\pm$0.08 magnitudes) on $H_R(40)$.  

Absolute magnitudes computed from Equation (\ref{abs}) are listed in Table (\ref{photometry}) and plotted in Figure (\ref{absmagplot}).  The error bars in the Figure are computed from $\sigma_T = (\sigma_m^2 + 0.08^2)^{1/2}$, where $\sigma_m$ is the measurement error listed in Table (\ref{photometry}) and 0.08 is our estimate of the phase function uncertainty. B1 shows an intrinsic brightening of $\sim$1 magnitude between the epoch of discovery in 2014 and perihelion in 2017.  A least-squares fit to the data from Table (\ref{photometry}), plotted in Figure (\ref{absmagplot}) as a red line, shows that the cross-section varies as $C_e \propto r_H^{-4.9\pm0.4}$.  This is  steeper than the $r_H^{-2}$ dependence expected of a super-volatile sublimator.  However, the gradient of this fit is heavily dependent on the one 2014 measurement and so we do not imbue it with further signifiance.

The absolute magnitude is related to the effective scattering cross-section, $C_e$, by 

\begin{equation}
C_e = \frac{1.5\times 10^6}{p_R} 10^{-0.4 H_R(40)}
\label{area}
\end{equation}

\noindent where $p_R$ is the red geometric  albedo.  Comet dust albedos are, in general, not well measured, with evidence for a wide range of values amongst comets and even temporal variability (declining from 0.12 to 0.04) within a single object (Ishiguro et al.~2010).  We here assume a nominal albedo $p_R$ = 0.1, while acknowledging that the true value may be larger or smaller by a factor of several, especially if the grains are icy (as, indeed, seems likely at 10 AU).  Derived values of $C_e$ from Equation (\ref{area}) are listed in Table (\ref{photometry}).  Both the large cross-sections listed there and the extended morphology of B1 indicate an origin by scattering from the coma, with a minimal (and poorly constrained) contribution from the central nucleus.   

The optical colors of B1 (B-V = 0.85$\pm$0.03, V-R = 0.58$\pm$0.03) are slightly redder than the average colors of 25 long-period comets (B-V = 0.78$\pm$0.02, V-R = 0.47$\pm$0.02, see Jewitt 2015), and redder than the Sun (for which B-V = 0.64$\pm$0.02, V-R = 0.35$\pm$0.01) but less red than the ``ultrared'' matter (B-V = 1.06$\pm$0.02, V-R = 0.66$\pm$0.02) that is a distinctive feature of many Kuiper belt objects (Jewitt 2015).  These colors are consistent with scattering from dust; the resonance fluorescence bands from abundant gas phase radicals (e.g.~CN, C$_2$, C$_3$) are heavily concentrated towards short wavelengths and would produce  colors bluer than observed, especially in B-V.  Indeed, cometary gas rarely dominates the optical spectrum except near $r_H \sim$ 1 AU, and has not been detected in a comet as far from the Sun as is B1.

\subsection{Structure of the Coma}
\begin{figure}[ht]
\centering
\includegraphics[width=0.95\textwidth]{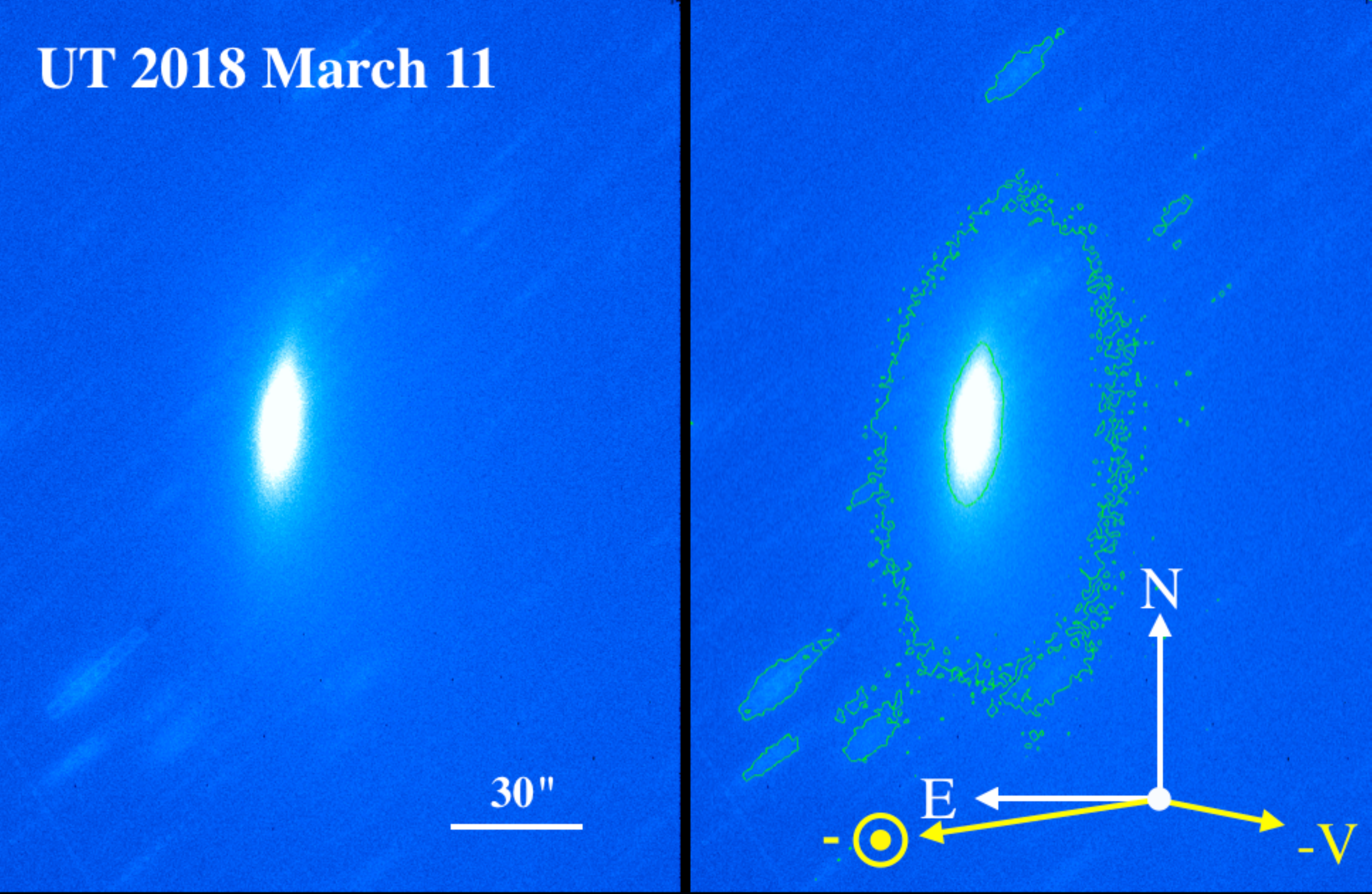}
\caption{(left): R-band image of C/2014 B1 taken UT 2018 March 11 showing the disk-like, roughly north-south elongation of the coma and the lack of any normal tail.   A scale bar shows  30\arcsec.  (right): Contoured version of the same image, with cardinal directions in white and the directions of the antisolar vector ($-{\odot}$) and the negative projected heliocentric velocity vector ($-V$)  in yellow.  The perpendicular extension of the coma is unique. \label{image} }
\end{figure}

\begin{figure}[ht]
\centering
\includegraphics[width=0.8\textwidth]{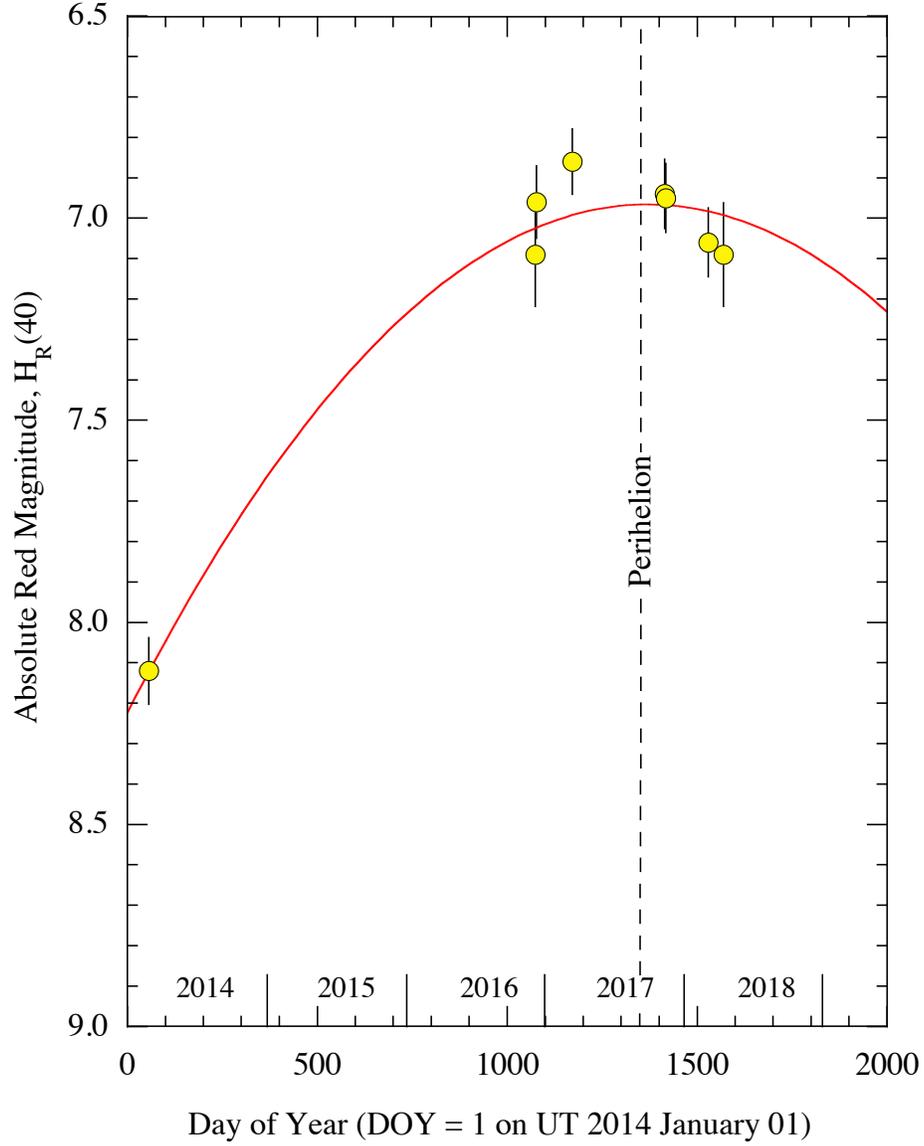}
\caption{Absolute magnitude within a 4$\times$10$^4$ km radius projected photometry aperture as a function of time, expressed as Day of Year.  The red curve is a best-fit in which the cross-section varies as  $C_e \propto r_H^{-4.9\pm0.4}$, to guide the eye.  The date of perihelion is indicated by the dashed vertical line.  \label{absmagplot} }
\end{figure}

While the comet intrinsically brightened by about one magnitude, the morphology of B1 in our data changed little from that in Figure (\ref{image}) in four years of observation.  In order to quantify this, we separately measured the position angles of the central axes of the coma extending to the north and the south of the nucleus, labeled $\theta_N$ and $\theta_S$, respectively, as a function of time.  To do this, we used perpendicular cross-sections at different distances from the nucleus.  The position angles were determined from linear least-squares fits to the position of peak surface brightness, weighted by the uncertainty on the position of the peak.  Table (\ref{photometry}) lists the resulting $\theta_N$ and $\theta_S$ together with their uncertainties.  For the latter, we found that the formal errors from the least-squares fits are small compared to systematic errors caused by structure in the sky background, especially at large projected distances from the nucleus where the surface brightness of the coma is very low.    (In addition, our representation of the long axis of the coma as linear  is itself only an approximation, and some of the images hint at curvature which we do not here further consider).  The listed errors are our best estimates of the true uncertainties in $\theta_N$ and $\theta_S$, including these systematic, background effects.  

\begin{figure}[ht]
\centering
\includegraphics[width=1.0\textwidth]{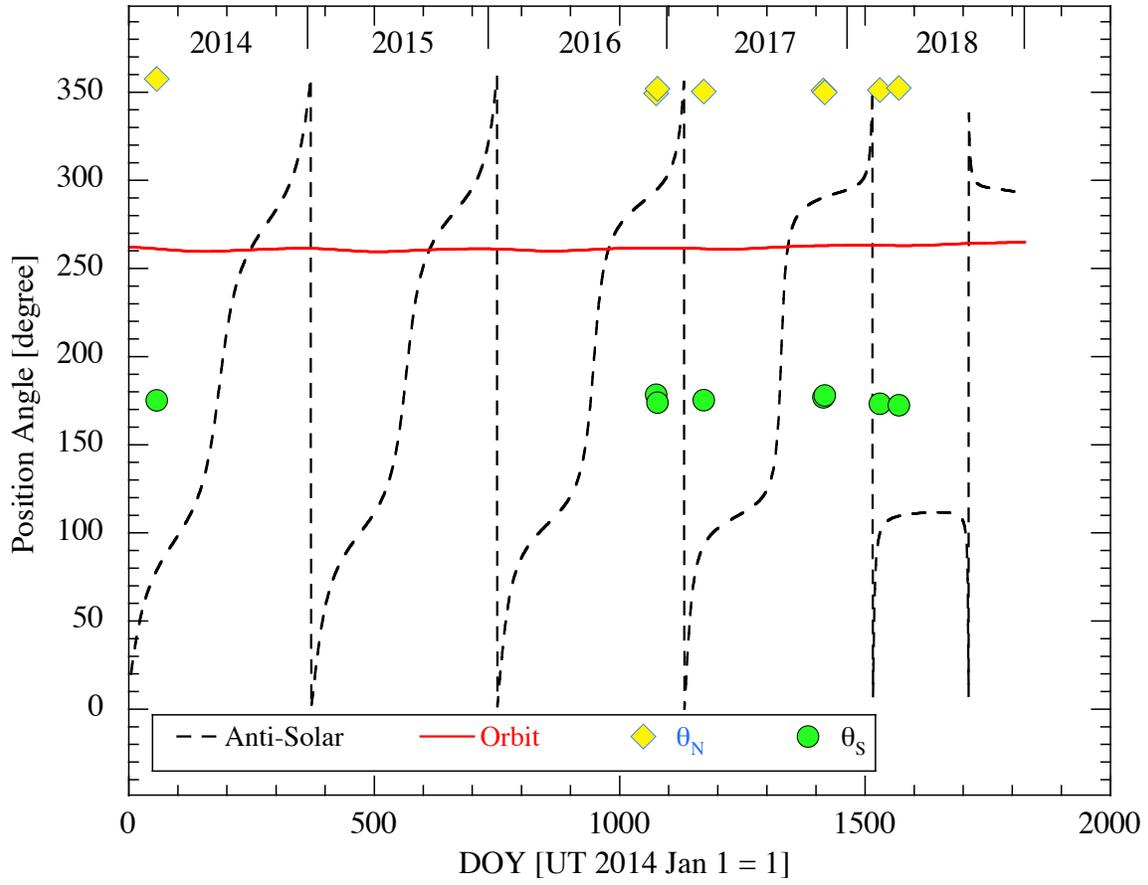}
\caption{Time dependence of the position angles of the North (yellow diamonds) and South (green circles) coma extensions from 2014 to 2018.  The solid red curve shows the position angle of the projected orbit.  The dashed black curve shows the position angle of the projected anti-solar direction.  \label{angle_plot} }
\end{figure}

Figure (\ref{angle_plot}) compares the measured angles,  as a function of time, with the position angles of the projected antisolar ($\theta_{-\odot}$) and orbital velocity ($\theta_{-V})$ vectors.   Neither $\theta_N$ nor $\theta_S$ shows any relation to  $\theta_{-\odot}$ or $\theta_{-V}$, showing that the shape of the coma is a property of the comet, not a result of external influences.

\begin{figure}[ht]
\centering
\includegraphics[width=1.0\textwidth]{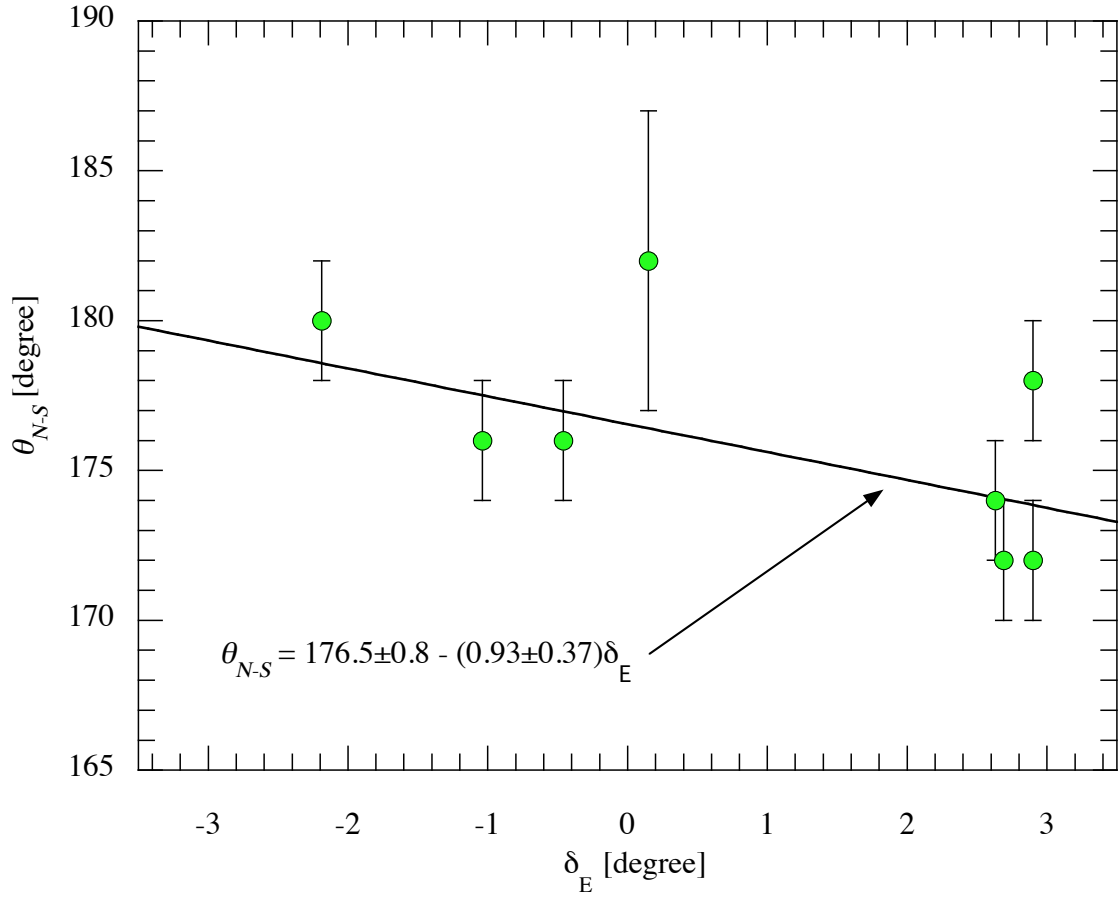}
\caption{Angle between the North and South arms of the coma vs.~the elevation of the Earth with respect to the orbital plane of B1.  The black line shows the weighted least-squares fit to the data.  \label{tilt} }
\end{figure}

The long axes of the coma measured to the north and the south of the nucleus appear nearly, but not quite, aligned.    We define the angle between the north and south extensions as $\theta_{N-S} = |\theta_N - \theta_S |$, finding a weighted mean value $\theta_{N-S}$ = 176\degr.1$\pm$0\degr.8 (Table \ref{photometry}), which is close to but significantly different from 180\degr.    The only hint of an angle relationship is between $\theta_{N-S}$ and the Earth elevation, $\delta_{E}$, shown in Figure (\ref{tilt})  (data from Tables \ref{geometry} and \ref{photometry}).  Such a relationship would be expected if $\theta_{N-S}$ is affected by projection of the three-dimensional dust distribution into the plane of the sky.  However, a weighted least-squares fit (plotted as a straight line in Figure \ref{tilt}) gives  $\theta_{N-S} = 176.5\pm0.8 - (0.93\pm0.37)\delta_{E}$, showing that the dependence  is significant  only at the 2.5$\sigma$ level.  We therefore do not interpret it further.

\begin{figure}[ht]
\centering
\includegraphics[width=1.0\textwidth]{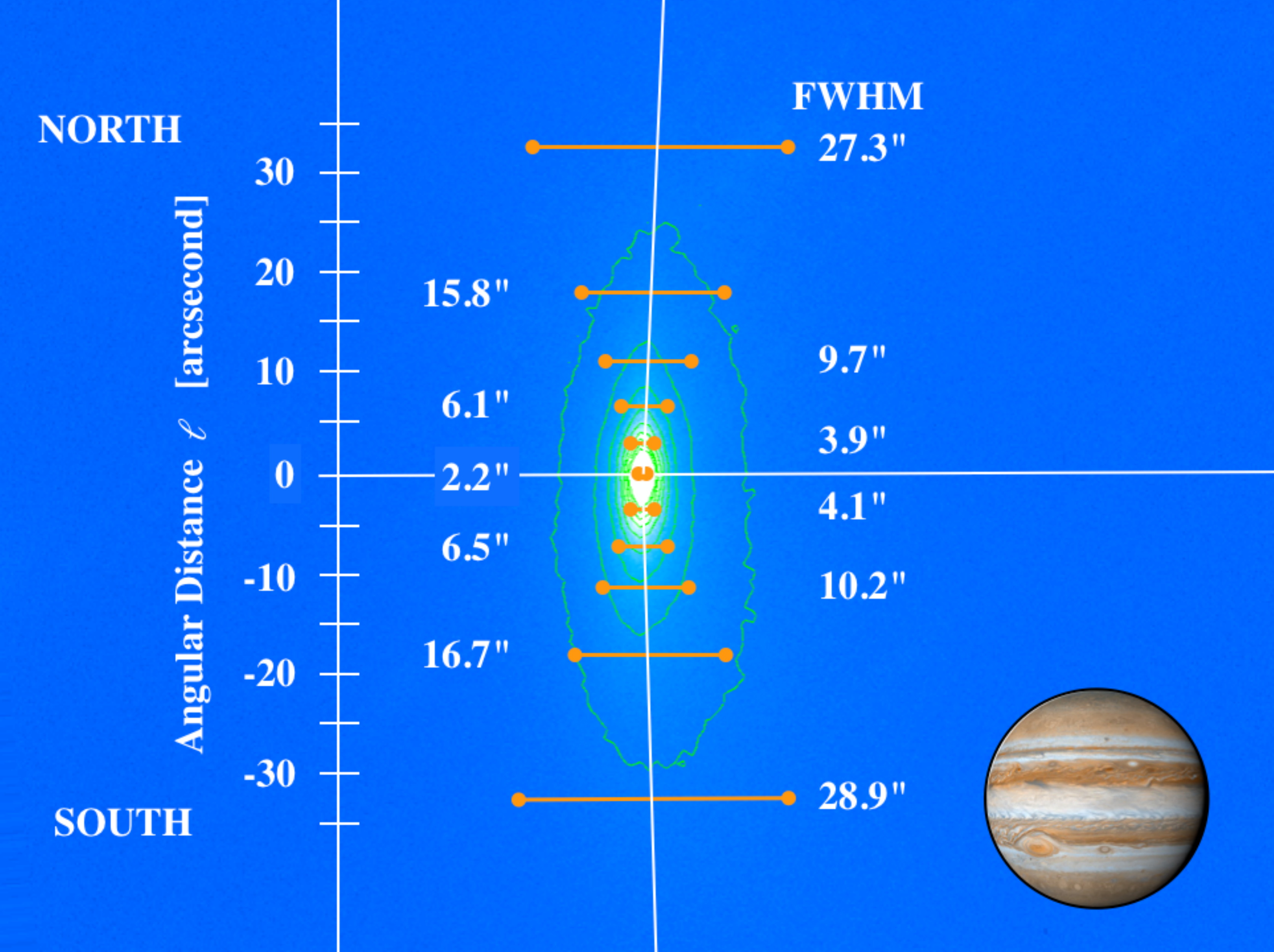}
\caption{A contoured image from UT 2018 March 11 is overlain by measurements of the full width at half maximum (orange bars, labelled by their numerical values) as a function of distance from the nucleus, $\ell$.  The inset image of Jupiter (140,000 km diameter) shows the scale.  \label{fwhm} }
\end{figure}

We also measured the  full width at half maximum  of the coma, as a function of $\ell$, the projected angular distance from the opto-center measured along the long axis of the coma.  To do this, we first rotated the image to bring the long axis to vertical, and then made horizontal cuts across the coma to determine the angular Full Width at Half Maximum (FWHM).  Figure (\ref{fwhm}) shows the resulting FWHM, called $\theta_{1/2}$ (orange bars), as a function of $\ell$.  As remarked earlier and apparent in Figure (\ref{image}), the optical appearance of B1 is discus-like, with a  width that decreases with increasing distance from the nucleus.  Figure (\ref{fwhm}) shows the FWHM in fact increases with distance from the nucleus, giving a double fan  type structure to the coma.    Least-squares fits to the data of Figure (\ref{fwhm})  having the form $\theta_{1/2} = a \ell + b$, give $a$ = 0.81$\pm$0.02 to the north and $a = -0.86\pm0.02$ to the south.  The opening angles of the fans are calculated from tan$^{-1} a$ = 39\degr$\pm$1\degr~in the north and 41\degr$\pm$1\degr~in the south.  The close similarity between these angles reflects the symmetry apparent in Figure (\ref{image}).   

\subsection{Radial Surface Brightness}
\begin{figure}[ht]
\centering
\includegraphics[width=0.9\textwidth]{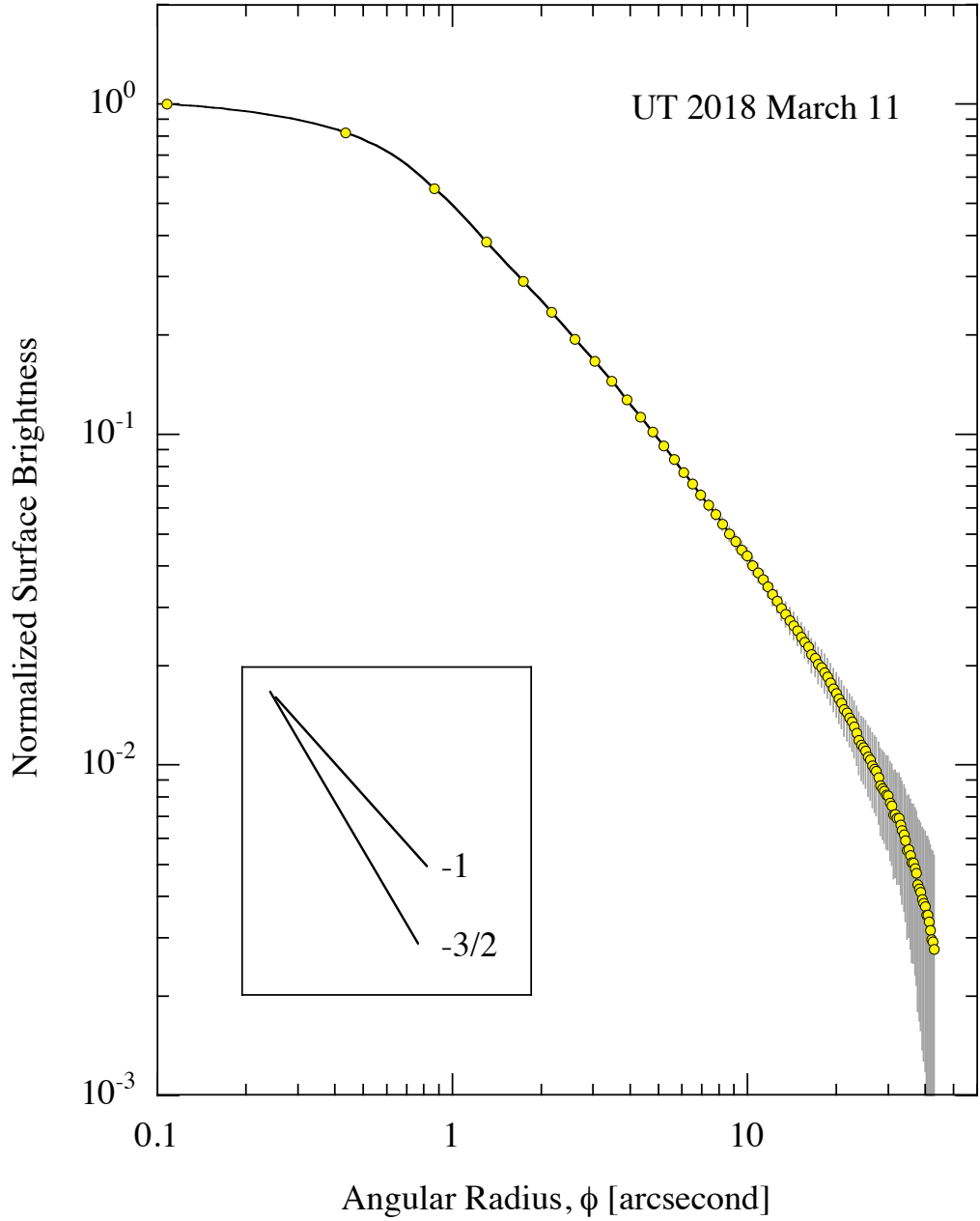}
\caption{Surface brightness profile of C/2014 B1 on UT 2018 March 11. The uncertainty, dominated by uncertainty in the sky background subtraction, is indicated by the grey shaded region.  \label{profile} }
\end{figure}

We computed the surface brightness profile, $\Sigma(\phi)$ [magnitudes per arcsecond$^{-2}$], where $\phi$ is the angular distance from the optocenter.  We present the data from UT 2018 March 11 because these have the least background contamination from field objects; profiles from other dates are similar but of lower quality.  The surface brightness within a nested set of concentric, circular apertures is plotted in Figure (\ref{profile}).  The background was determined from the median of the pixels in a concentric annulus having inner and outer radii 54\arcsec~and 64\arcsec, respectively.   At small angles from the optocenter, $\phi \lesssim 2\arcsec$, the surface brightness is affected by convolution with the point spread function of the image.  Conversely, at large angles, $\phi \gtrsim$ 10\arcsec, the accuracy of the surface brightness determination is increasingly limited by the uncertainty in the brightness of the background sky.  We estimated the sky uncertainty by experimenting with different sky apertures; its value is dominated by imperfections in the flat fielding of the data and by residual trails of field objects.

The  surface brightness profile is  usefully characterized by its gradient, $m = d \ln \Sigma(\phi)/d \ln \phi$.  A coma produced in steady-state should, in the absence of radiation pressure, have $m = -1$ while, in the presence of radiation pressure,  the gradient should asymptotically steepen towards $m = -1.5$ as a result of the steady acceleration of the particles (Jewitt and Meech 1987).   While most simply derived for the case of a spherically symmetric, steady-state coma, the $m$ = -1 gradient is a consequence only of the equation of continuity, and applies equally well to anisotropic comae provided they remain optically thin.  Therefore, it is reasonable to use concentric apertures and to interpret the  surface brightness gradient of B1 in the above terms even though its coma is clearly anisotropic.  The measured gradient in our best data (from UT 2018 March  11) B1 is $m = -1.10\pm0.01$ in the range $2\arcsec \le \phi \le 10\arcsec$, more similar to  the steady-state value than to the radiation pressure limit.  However, we also find that $m$ is itself an apparent function of $\phi$ ($m$ becomes more negative as $\phi$ increases, see Figure \ref{profile}).  If real, this steepening coma is probably due either to fading of the grains (e.g.~due to ice loss or to their disaggregation into tiny sub-components) or to radiation pressure acceleration at large angles.  However, the outer portion of the surface brightness profile where the gradient changes is precisely where the sky subtraction uncertainties become dominant (see error bars in Figure \ref{profile}) and so we cannot reach any definitive interpretation of this region.   Gradients measured in other data from Table (\ref{geometry}) are consistent with the value measured on March 11, but are less accurate as a result of background uncertainties.

\section{DISCUSSION}

The following special features  of B1 are deserving of explanation;

\noindent 1) a persistent bi-convex lens shaped appearance (Figure \ref{image}) with an underlying  cone-like dust distribution (Figure \ref{fwhm}). 

\noindent 2)  the weak evidence for the effect of radiation pressure on the dust (Figures \ref{image} and \ref{profile})

\noindent 3) the long-term stability of the position angle of the coma (Figure \ref{angle_plot})

\noindent 4)  the fact that the position angles of the coma to the north and the south of the nucleus differ by nearly (but not exactly) 180\degr~(Table \ref{photometry})

Before resorting to a multi-parameter numerical model, we first consider what more can be learned about B1 based on the above observations and order of magnitude considerations.

\subsection{Particle Size}
We can immediately exclude the possibility that small particles dominate the scattering cross-section in the coma of B1 because these should be deflected into a classical, radiation pressure swept tail that is not seen.   The outer isophotes (contoured in the right-hand panel of Figure \ref{image}) do show a large-scale ($\gtrsim$ 30\arcsec) east-west asymmetry that might be related to radiation pressure, corresponding to a distance $L \sim 2.2\times 10^8$ m at the comet, in the plane of the sky.  If this ``tail'' makes an angle $\alpha$ to the line of sight, the physical length is $L/\tan(\alpha)$.  We take $\alpha$ = 4\degr~(equal to the phase angle, see Table \ref{geometry}) to find $L \sim 3\times 10^9$ m, as our best estimate of the scale on which radiation pressure acts.  

Following convention, we define $\beta$ as the ratio of the acceleration due to radiation pressure to the acceleration due to solar gravity.  For a sphere of radius $a$ and density $\rho$ this ratio is

\begin{equation}
\beta = \frac{3Q_{pr} L_{\odot}}{16\pi G  M_{\odot}c  \rho a}
\label{beta}
\end{equation}

\noindent where $L_{\odot} = 4\times 10^{26}$ W and $M_{\odot} = 2\times 10^{30}$ kg are the luminosity and mass of the Sun, $G = 6.67\times 10^{-11}$ N kg$^{-2}$ m$^2$ is the gravitational constant and $c = 3\times 10^8$ m s$^{-1}$ is the speed of light.  Dimensionless quantity $Q_{pr}$ is the radiation pressure efficiency, which is of order unity for particles with $a \gtrsim \lambda$, where $\lambda \sim$ 0.5 $\mu$m is the wavelength of light, and $Q_{pr} \ll 1$ otherwise (Bohren and Huffman 1983).

Substituting $\rho$ = 500 kg m$^{-3}$, Equation (\ref{beta}) gives  $\beta \sim a_{\mu m}^{-1}$, where $a_{\mu m}$ is the particle radius expressed in microns. The distance over which a particle is accelerated in time, $t$, is just $L =  g_{\odot}(1) t^2/(a_{\mu m}r_H^2)$, where $g_{\odot}(1)$ = 0.006 m s$^{-2}$ is the gravitational acceleration at 1 AU and $r_H $ is expressed in AU.  We take $r_H $ = 10 AU as representative for the period of observations discussed here.  For example, consider a micron-sized particle ($a_{\mu m}$ = 1) released from the nucleus with zero initial velocity.  B1 has been active for $t >$ 4 years (1.2$\times 10^8$ s, c.f.~Table \ref{photometry}), during which time the distance travelled relative to the nucleus would be $L \sim 4\times 10^{11}$ m ($\sim$ 2.5 AU), fully two orders of magnitude larger than the scale of the coma in Figure (\ref{image}).  Even given the crude nature of this calculation, it is evident that the absence of a  tail in B1 requires that the optically dominant particles have sizes $a \gg$ 1 $\mu$m. 

If we instead keep $t = 1.2\times 10^8$ s and set $L = 3\times 10^{9}$ m, representing the approximate scale for radiation pressure deflection as estimated above, we solve to find  $\beta \lesssim$ 0.01 ($a \gtrsim 100~\mu$m) for the effective size of the particles.  Smaller particles ($\beta \gtrsim$ 0.01) would have been accelerated by radiation pressure over distances larger than the size of the observed coma, forming a tail that is not observed.   This argument is clearly  approximate, but it serves to make the point that the stability of the morphology over four years and the absence of a radiation-swept tail imply that the effective particle size in B1 is very large.  The prevalence of large particles in other distant comets has been noted (e.g.~Jewitt et al.~2017) and interpreted as a reflection of the effects of inter-particle cohesion, which binds small particles to the nucleus (Jewitt et al.~2018).

\subsection{Production Rate}

The dust production rate  can be  estimated to order of magnitude from the time dependence of the scattering cross-section in Table (\ref{photometry}).  We observe that between UT 2014 February 26 and 2017 March 22, an interval of $\Delta t = 9.6\times 10^7$ s, the cross-section increased by $\Delta C_e = 1.8\times 10^4$ km$^2$.  For an optically thin collection of spheres, the rates of change of the mass and cross-section are related by

\begin{equation}
\frac{dM}{dt} = \frac{4}{3} \rho \overline{a} \frac{\Delta C_e}{\Delta t}
\label{dmbdt}
\end{equation}

\noindent where $\rho$ = 500 kg m$^{-3}$ is the assumed grain density (e.g.~see Figure 2 of Fulle et al.~2018) and $\overline{a}$ is the average particle radius.  With $\overline{a} \gtrsim$ 100 $\mu$m, we find from Equation (\ref{dmbdt}) that $dM/dt \gtrsim$ 10 kg s$^{-1}$.  Strictly, $dM/dt$ represents the difference between mass added to the 40,000 km radius photometry aperture at the nucleus and lost from the aperture by outflow at its outer edge.  It therefore sets a strong lower limit to the mass loss rate from the nucleus.

B1 is too distant and too cold for water ice to sublimate. However, its heliocentric distance and temperature are such that  exposed amorphous water ice, if present, should crystallize (Guilbert-Lepoutre 2012).  Crystallization, accompanied by the release of trapped gases and the expulsion of dust through gas drag forces could account for, or perhaps trigger, the observed activity.  B1 is also warm enough for a very small area of exposed CO or CO$_2$ ice to drive the mass loss.  By solving the energy balance equation (e.g.~as in Jewitt et al.~2017, 2018) for sublimating ice at the sub-solar point on a nucleus at $r_H$ = 10 AU, we find  maximum CO and CO$_2$ sublimation rates of $f_s(CO) = 4\times 10^{-5}$ kg m$^{-2}$ s$^{-1}$ and $f_s(CO_2) = 2\times 10^{-5}$ kg m$^{-2}$ s$^{-1}$, respectively.  To supply 10 kg s$^{-1}$ would require an ice patch of 0.3 km$^2$ for CO and 0.6 km$^2$ for CO$_2$.  These are very modest surface areas, as may be seen, for example, by comparison with the Centaur comet 29P/Schwassman-Wachmann 1. At $r_H$ = 6 AU, 29P loses CO at $\sim$2000 kg s$^{-1}$ (Senay and Jewitt 1994), requiring a sublimating surface of 16 km$^2$, calculated in the same way.  If the dust to gas mass ratio, $f_{dg}$, is different from unity then these sublimating areas should be multiplied by $f_{dg}^{-1}$.  The ratio was $f_{dg}$ = 4 in 67P/Churyumov-Gerasimenko (Rotundi et al.~2015), $f_{dg} >$ 5 in long-period comet C/1995 O1 (Hale-Bopp) (Jewitt and Matthews 1999) and values as high as $f_{dg}$ = 30 have been reported (Reach et al.~2000).

\subsection{Disk Ejection Model}
\label{disky}
The images suggest that the coma of B1 is a disk-shaped figure viewed edge-wise.  If we identify the long axis of the coma with the projected rotational equator, then  the projected pole of the nucleus must lie near the plane of the sky  at position angle $\sim$80\degr~to 85\degr~in Figure (\ref{image}).   The expanding north and south arms of the coma (Figure \ref{fwhm}) then represent the equatorial plane of the coma, where more dust is emitted than at larger latitudes.   In this scenario, the near north-south symmetry of the coma is a product of rotation.   Physically, we imagine that such a disk could result only  if the  equatorial rotational velocity of the nucleus is a significant fraction of the gravitational escape velocity and of the dust launch speeds.  Particles would escape preferentially from low latitudes with the assistance of centripetal acceleration, while those emanating at higher latitudes would fall back along suborbital trajectories without contributing to the coma.    Depending on the ratio of the rotation period to the night-side cooling time, the dust ejection could be continuous in azimuth or restricted to the day-side of the nucleus.  We envision that the angle $|\theta_N - \theta_S|$ = 176\degr.1$\pm$0\degr.4 is not quite 180\degr~because the Earth is not exactly in the projected equator of B1.

To explore this geometry and to test some of the inferences made above based on order-of-magnitude considerations, we used the Monte Carlo dust dynamics model of Ishiguro et al.~(2007).  The model follows the motions of dust particles under the action of solar gravity and radiation pressure, after their ejection from the nucleus according to specified speed and direction parameters.  With many free or under-constrained parameters the model cannot, in  general, offer unique solutions for the particle and ejection parameters. The Ishiguro model is nevertheless valuable in allowing an exploration of parameter regimes that are consistent with the imaging data.  

We explored a range of parameters to try to match the morphology of B1 on UT 2018 March 11.  As expected, plausible solutions from the generated models all required the optical dominance of large particles ejected from the nucleus with small speeds.  The models also required emission from the nucleus over a narrow range of latitudes and a pole direction inclined to the line of sight by $\sim$90\degr.  In all these regards, the Monte Carlo models support the inferences made above on the basis of simple physical arguments.  

\begin{figure}[ht]
\centering
\includegraphics[width=0.9\textwidth]{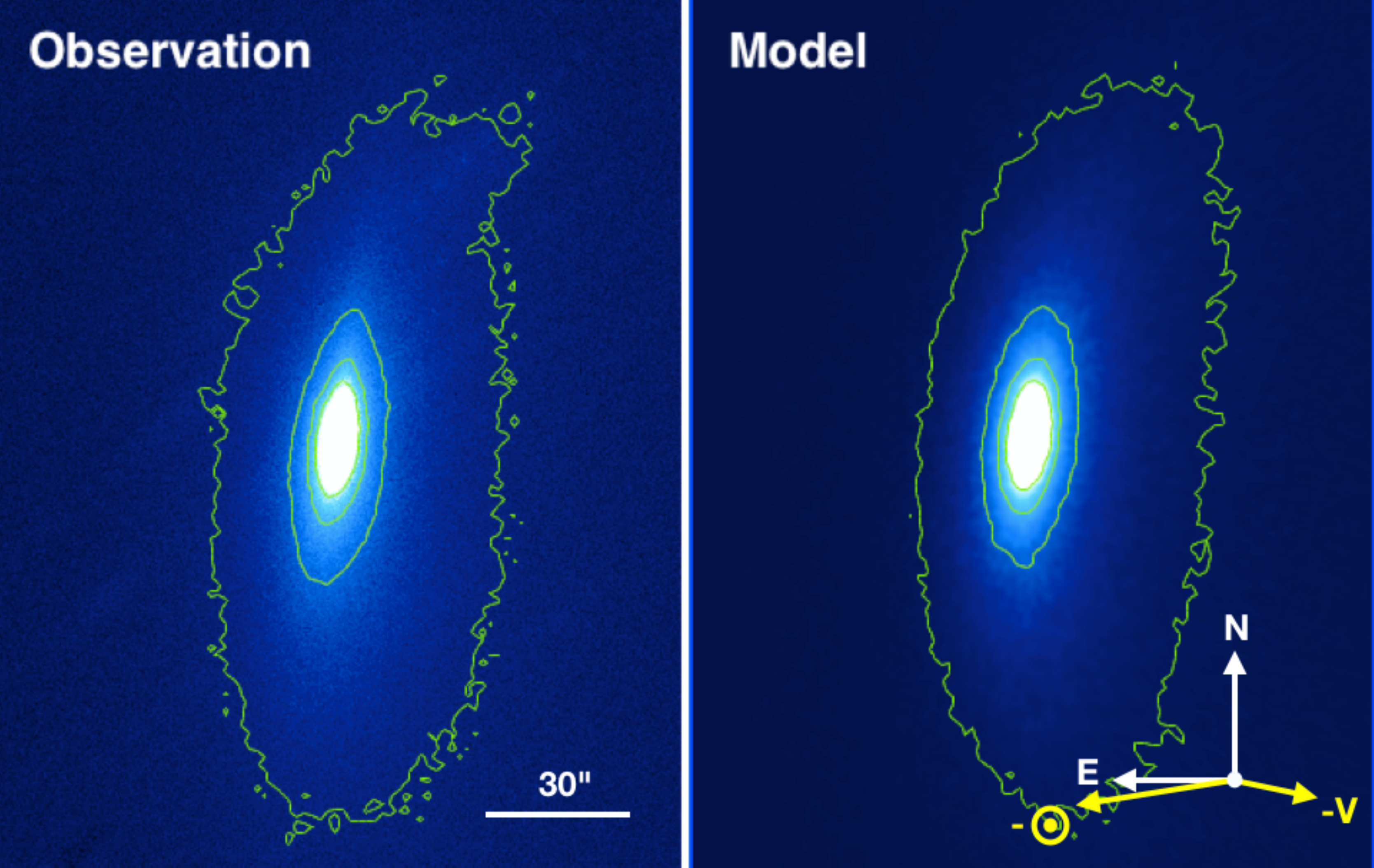}
\caption{(left) UT 2018 March 11 image compared with (right) best-fit Monte Carlo model with parameters described in Section (\ref{disky}).  The distortion of the upper right portion of the outer isophote in the data is caused by imperfect removal of a trailed field galaxy.  \label{bestmodel} }
\end{figure}

\begin{figure}[ht]
\centering
\includegraphics[width=0.9\textwidth]{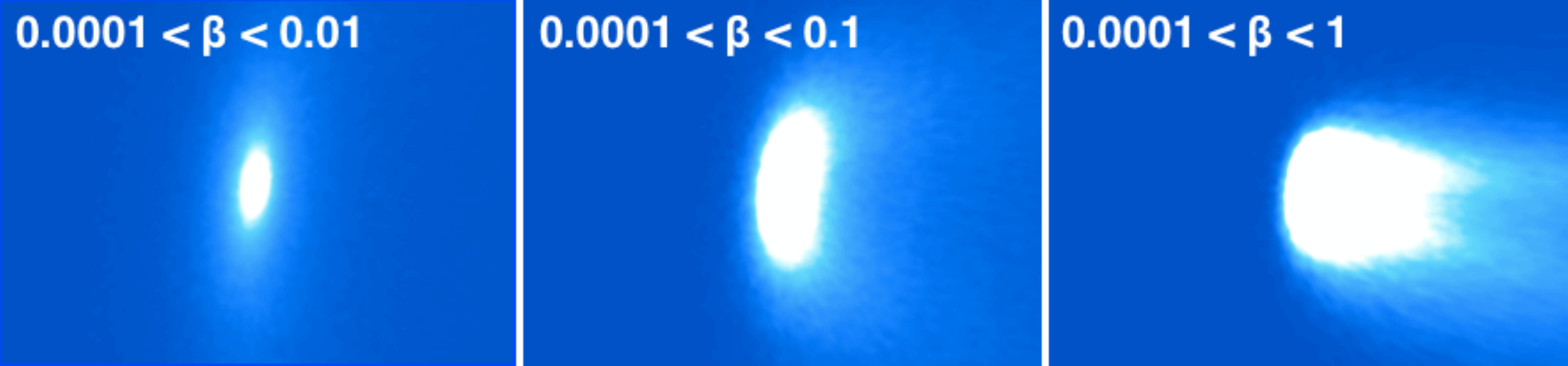}
\caption{Effect of $\beta_{max}$ on the morphology, corresponding to minimum particle radii 100 $\mu$m, 10 $\mu$m and 1 $\mu$m, from left to right. Small particles occupy a radiation-pressure swept tail to the west (right) that is not present in B1. This simulation is for UT 2018 March 11, for which the anti-solar and negative heliocentric velocity vectors are as indicated in Figure (\ref{bestmodel}). \label{betamax} }
\end{figure}

A specific example is shown in Figure (\ref{bestmodel}).  The parameter assumptions used to generate this model include a power-law distribution of particle sizes, with differential index $q$ = 3.5, dust emission continuous over four years, a velocity vs.~radiation pressure parameter relation, $V = V_1 \beta^{u1}$, with $V_1$ = 100 m s$^{-1}$ and $u1$ = 1/2, and a range of particle sizes from $\beta_{min}$ = 10$^{-4}$ to $\beta_{max} = 10^{-2}$.  Experiments show that, while $\beta_{min}$ is poorly defined, $\beta_{max}$ (a measure of the smallest particles) cannot be substantially increased without destroying the similarity between the model and the data through the formation of a prominent tail (Figure \ref{betamax}).  Thus, we conclude that the particle radii fall in the range from $a_{min}$ = 0.1 mm to $a_{max} \sim$ 10 mm and that their speeds of ejection, by the above relation, lie in the range $1 \lesssim V \le $ 10 m s$^{-1}$.  The best-fit rotation vector points at right ascension 90\degr~and declination -3\degr~(or its opposite).  The position angle of the rotation vector, projected into the plane of the sky, is 84\degr, in substantial agreement with the visual estimate of 80\degr~to 85\degr, mentioned above.    In the model shown, dust is ejected from a band extending $\pm$10\degr~from the equator.  Isotropic and hemispheric emission models failed to reproduce the lens-like appearance of B1 and can be rejected.

In short, the Monte Carlo models substantially support inferences made above on the basis of visual examination of B1.  The peculiar and stable discus shape of the coma can be matched by equatorial emission of large ($a >$ 0.1 mm), slow ($V <$ 10 m s$^{-1}$) particles from a nucleus having a spin pole near the plane of the sky.

\subsection{Opposing Jets Model}

An alternate possibility is  that the north and south arms of the coma reflect  ejection from  diametrically opposite  active area (``jet'') sources on the nucleus, one for each of the two arms.  We consider this possibility less attractive than the disk ejection model for several reasons.  First, the assumption of two comparably active, diametrically opposite jets continuously active for $\ge$ 4 years, while possible, is  completely ad-hoc. Second, this conjecture is energetically problematic because the source regions would necessarily be located about $\pm$90\degr~from the subsolar point, near the sunrise and sunset terminators, where the illumination needed to drive the mass loss  is weakest.  Third, jets in comets are typically curved and/or temporally modulated by nucleus rotation (e.g.~Larson et al.~1987), whereas the coma in B1 is not.   

\subsection{Lorentz Force Model}

The fact that the elongation of B1 is unrelated to the anti-solar and projected orbit directions raises the possibility that other forces (in particular, the Lorentz force), might be at play.  The Lorentz force has been suggested as an agent in shaping sub-structures (the ``striae'')  in the tails of some near-Sun comets (Ip et al.~1985) and in the coma of distant comet C/1995 O1 (Hale-Bopp)  (Kramer at al.~2014) but has not otherwise been suspected.   

Dust particles exposed to sunlight are charged to a positive potential $W \sim$ 5 V by the loss of photoelectrons (Kimura and Mann~1998, Pavlu et al.~2008).   On a grain of radius $a$, this potential corresponds to a charge $q = 4\pi \epsilon_0 W a$, where $\varepsilon_0 = 8.85\times10^{-12}$ F m$^{-1}$ is the permittivity of free space.  The charge on the grain renders the particle susceptible to the  Lorentz force, $\vec{F_L}$,  given by $\vec{F_L} = q(\vec{v}\times \vec{B})$, where  $\vec{B}$ is the magnetic flux density in the solar wind and $\vec{v}$ is the relative velocity of the wind passing the comet.   Since $\vec{v}$ is radial, the relevant component of the magnetic field is the azimuthal component, and $\vec{F_L}$ is perpendicular to the solar system mid-plane.  In-situ measurements over a range of heliocentric distances   show that $\vec{B}$, while highly variable on the $\sim$1 month solar rotation timescale, varies  inversely with $r_H$. We write $B = B_1 r_H^{-1}$  where $B_1 = 600$ (T m) is a constant (determined from Figure 11 of Balogh and Erdos 2013) and $r_H$ is in meters.  For example, the flux density at 10 AU is $B = 0.4$ nT, with factor-of-two fluctuations occurring on the solar rotation timescale.  The solar wind speed is also variable, depending on activity on the Sun, but is well represented by  $v$ = 500 km s$^{-1}$, roughly independent of $r_H$ (Balogh and Erdos 2013).  

Combining these relations we define the ratio of the  Lorentz force, $|\vec{F_L}|$, to the gravitational force as  the ``magnetic $\beta$'', $\beta_m$;

\begin{equation}
\beta_m = \frac{3 \varepsilon_0 B_1 W r_H v }{G M_{\odot} \rho a^2}
\label{beta_m}
\end{equation}

\noindent where $r_H$ is expressed in m, and the other quantities are as defined above.  

Finally, the ratio of the Lorentz acceleration to the radiation pressure acceleration from Equations (\ref{beta}) and (\ref{beta_m}) is 

\begin{equation}
\frac{\beta_m}{\beta} = \frac{16\pi \epsilon_0 B_1 W  v c }{Q_{pr} L_{\odot}} \left[ \frac{r_H}{a}\right]
\end{equation}

\noindent Substituting, and now expressing $r_H$ in AU and $a$ in microns, we obtain

\begin{equation}
\frac{\beta_m}{\beta} = 0.07 \left[\frac{r_H}{a_{\mu m}}\right].
\label{betaratio}
\end{equation}

\begin{figure}[ht]
\centering
\includegraphics[width=0.9\textwidth]{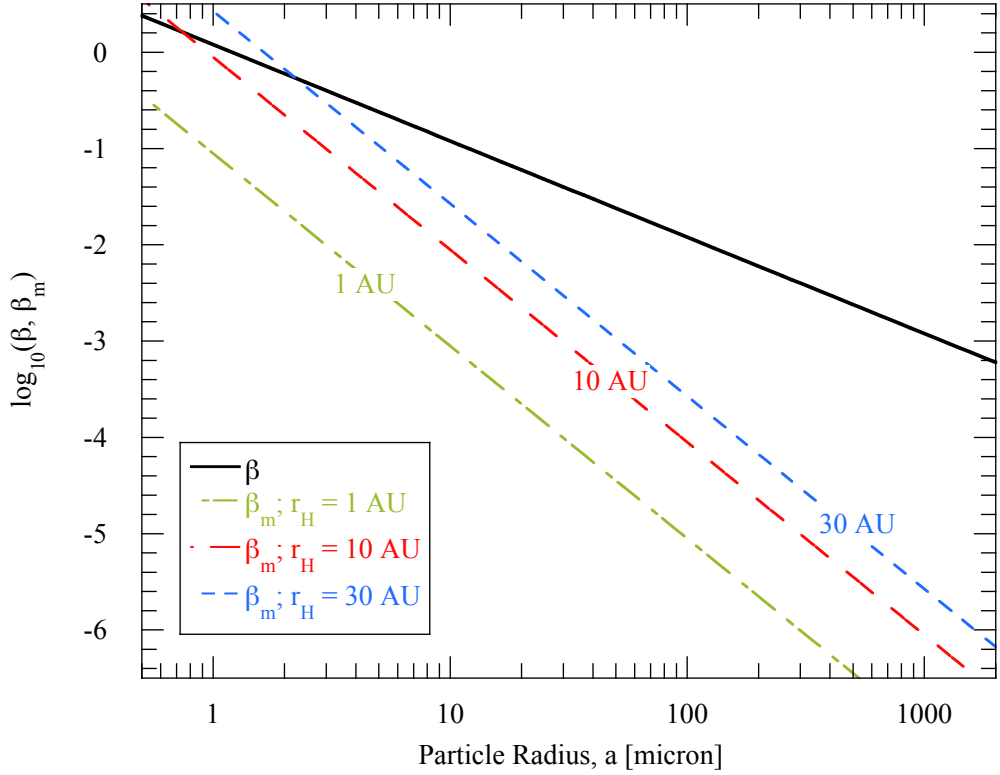}
\caption{Radiation pressure $\beta$ (solid black line, Equation \ref{beta}) and magnetic $\beta_m$ (dashed colored lines, Equation \ref{beta_m}) vs.~particle radius.  $\beta_m$ is plotted for three representative values of the heliocentric distance, $r_H$, as labeled. \label{betaplot} }
\end{figure}

\noindent Equations (\ref{beta}) and (\ref{beta_m}) are plotted in Figure (\ref{betaplot}), for three values of the heliocentric distance, $r_H$ = 1, 10 and 30 AU.  The Figure and Equation (\ref{betaratio}) show that, for example, at $r_H$ = 1 AU, $\beta = \beta_m$ for $a = 0.07~\mu$m.  Such tiny particles ($a/\lambda \sim 0.1$) are inefficient optical scatterers and, therefore, magnetic effects are not observed in comets near Earth even though tiny particles are abundant (McDonnell et al.~1986).  At the perihelion distance of B1, namely $r_H \sim$ 10 AU, we find from Equation (\ref{betaratio}) that $\beta = \beta_m$ when $a = 0.7~\mu$m.  Such micron-sized particles are efficient scatterers of optical photons and would be detected if they were abundant in B1.  As described above, small particles are depleted from the coma of B1 as a result of cohesion and the effective particle radius  is instead a relatively large $a \gtrsim 100~\mu$m.  Substituting $r_H$ = 10 AU, $a_{\mu m} \gtrsim$ 100 into Equation (\ref{betaratio}) we find   $\beta_m/\beta \lesssim  0.007$ showing that the effects of the Lorentz force can  be safely ignored in the context of the lens-like morphology of this comet.  

The images of B1 also provide other evidence against the role of Lorentz forces.  This is because, with $\vec{v}$ radial and $\vec{B}$ acting in the azimuthal direction,  the direction of $\vec{F_L}$ is necessarily close to the normal to the ecliptic, at all times.   Between 2014 and 2018, the ecliptic longitude of B1 changed by $\sim$60\degr, and the position angle of the ecliptic normal varied from 0\degr~to 20\degr.  In contrast,  the measured position angles of the north and south arms of the coma are stable to within the measurement uncertainties (mostly $\pm$1\degr~or $\pm$2\degr) in the interval 2014 - 2018, and certainly do not show any variation as large as 20\degr~(Table \ref{photometry}).   In addition, we note that the optical (BVR) colors of B1 are redder than sunlight, typical of all reliably measured comets (Jewitt 2015), and so provide no evidence for the  optically small (blue) particles that would be affected by Lorentz forces.

\subsection{Concluding Remarks}

The special morphology of B1 is an artifact of several effects.  First, the dust particles must be ejected slowly, with a speed, $V$, comparable to the gravitational escape velocity from the nucleus, $V_e$. We write $V_e \sim 0.5 r_n$ (m s$^{-1}$), where $r_n$ is the nucleus radius expressed in kilometers and where we have assumed a nucleus density $\rho$ = 500 kg m$^{-3}$ (c.f.~Jorda et al.~2016, Kokotanekova et al.~2017).  We set $V = V_e$.  Then, the speed range $1 \lesssim V \le $ 10 m s$^{-1}$ corresponds to nucleus radii 2 $\le r_n \le 20$ km.    While the radius of the nucleus of B1 has not been measured, values in this range are typical of well-measured cometary nuclei (Lamy et al.~2004). Second, the equatorial rotational velocity of the nucleus must be a significant fraction of $V_e$, in order to provide modulation of the ejection by latitude.  For a sphere of density $\rho$, the critical period is $P = (3\pi/(G \rho))^{1/2}$ which, with $\rho$ = 500 kg m$^{-3}$ gives $P = 1.2\times 10^4$ s (3.3 hr).  Somewhat longer critical rotation periods are possible for aspherical nuclei in rotation about a short axis.  Third, we suppose that the Earth is located close to the projected rotational equator of the nucleus of B1.   Our proposed model of B1 can thus be tested for consistency by determination of the nucleus size and by measurement of its rotation vector.  Such measurements will require  the use of high resolution imaging with the Hubble Space Telescope; our proposal to obtain such images was recently rejected.

The time-integrated mass loss from B1 is not well constrained.  To take an extreme case, we note that the rate at perihelion is $dM/dt \gtrsim$ 10 kg s$^{-1}$ and suppose that this rate has been sustained at all distances $r_H \le$ 30 AU.  On its current orbit, B1 spends 30 years (10$^9$ s) with $r_H \le$ 30 AU, corresponding to a total mass loss of $\sim$10$^{10}$ kg.  This compares with the mass of a 2 km to 20 km radius nucleus $M \sim$ 10$^{13}$ kg to 10$^{16}$ kg.  A fractional mass loss of $\Delta M/M \sim$ 10$^{-3}$ to 10$^{-6}$ is unlikely to generate a measurable non-gravitational acceleration, or to materially change the spin rate of the nucleus.  Moreover, while B1 presumably made previous journeys through the planetary region, it seems unlikely that prior mass loss has rivalled a significant fraction of the nucleus mass, or that outgassing torques have changed the spin of the nucleus.  Instead, we imagine that the  nucleus spin is  primordial, accumulated as the nucleus grew,  or in some collisional event occurring in the ancient protoplanetary disk.  Whatever the origin of the spin, mass loss has been limited to those brief, near-perihelion periods when gas drag forces exceed gravitational and cohesive forces binding material to the nucleus.

Finally, we ask why has the discus morphology  not been noticed in other comets?  The probable reason is that most  observations in the published literature refer to comets much closer to the Sun than B1, usually with $r_H \le$ 5 AU and frequently with $r_H$ as small as 1 to 2 AU.  In these comets, the equilibrium temperatures are higher and the sublimation fluxes (which depend exponentially on temperature) are orders of magnitude larger than in B1.  As a result, the cohesion bottleneck (in which  small particles are bound to the nucleus at large $r_H$), is broken, and small particles flood the coma (Gundlach et al.~2015, Jewitt et al.~2018).  Because they are small, the ejected particles in near-Sun comets attain terminal velocities $V \gg V_e$, erasing the role of centripetal assistance.  While  the discus morphology is not expected (and has not been reported) in comets nearer the Sun, we  predict that discus comets will become more common as future studies increasingly probe  comets far from the Sun, particularly those discovered pre-perihelion and beyond the cohesion bottleneck (Jewitt et al.~2018).

\clearpage

\section{SUMMARY}

We present a  study of the high-perihelion, long-period comet C/2014 B1 (Schwartz)  in the heliocentric distance range 11.9 AU to 9.6 AU.

\begin{enumerate}

\item The comet exhibits a unique and morphologically stable discus-shaped coma, whose orientation remains fixed even as the projected anti-solar and negative velocity vectors vary in response to the changing observing geometry.   This shows that the coma morphology is an intrinsic property of the comet, not a product of external influences. 

\item The lack of a prominent dust tail implies a large mean particle radius, 0.1 $\le a \le$ 10 mm, too large to be significantly affected by radiation pressure. The depletion of smaller particles is tentatively attributed to  inter-particle cohesion, whose effects dominate at small sizes.   Even larger particles may be present, but are rare.

\item  The absolute brightness of the coma increases by $\sim$1 magnitude, proving that mass-loss is on-going at 10 AU.
 We infer a dust mass production rate, $dM/dt \gtrsim$ 10 kg s$^{-1}$, that can be sustained by equilibrium sublimation from exposed supervolatile ices covering as little as 0.3 km$^2$ (for CO) to 0.6 km$^2$ (for CO$_2$) of the surface.  Water ice is too cold to sublimate at 10 AU, but crystallization of amorphous water ice might play a role in liberating trapped supervolatiles.

\item The discus-shaped coma is consistent with preferential equatorial ejection of dust from a nucleus whose rotational pole lies close to the plane of the sky, and from which the ejection speed (1-10 m/s) is comparable to the escape speed.  Centripetal assistance near the equator can then help launch large dust particles that cannot escape from high latitudes.   A Monte Carlo simulation based on this scenario successfully matches the morphology.  The implied nucleus radius is 2 - 20 km.

\item Based on our model, we predict that the discus shape should be most common in cometary comae that are dominated by large, slow particles, since only these would give rise to a latitude modulation of the dust ejection. The effect is absent in near-Sun comets, because the strong gas flux launches smaller particles far above the escape speed.
\end{enumerate}

\acknowledgments
Based in part on observations made with the Nordic Optical Telescope, operated by the Nordic Optical Telescope Scientific Association at the Observatorio del Roque de los Muchachos, La Palma, Spain, of the Instituto de Astrofisica de Canarias.  Some of the data  were obtained at the W. M. Keck Observatory, which is operated as a scientific partnership among the California Institute of Technology, the University of California and the National Aeronautics and Space Administration. The Observatory was made possible by the generous financial support of the W. M. Keck Foundation.  Also based in part on data from Kitt Peak National Observatory, National Optical Astronomy Observatory, which is operated by the Association of Universities for Research in Astronomy (AURA) under a cooperative agreement with the National Science Foundation. We thank Jessica Agarwal, Jing Li, Man-To Hui, Quanzhi Ye and the anonymous referee for comments on the manuscript.


%% in parentheses, after the keyword, but they will not be verified.

%{\it Facilities:}  \facility{}


\clearpage

\begin{deluxetable}{llcllllcrrrrr}
\tabletypesize{\scriptsize}
%\rotate
\tablecaption{Observing Geometry 
\label{geometry}}
\tablewidth{0pt}
\tablehead{\colhead{UT Date} & \colhead{UT} &\colhead{DOY} & Tel\tablenotemark{a} & Camera\tablenotemark{b} &  FOV\tablenotemark{c}  & Scale\tablenotemark{d} &  \colhead{$r_H$\tablenotemark{e}} & \colhead{$\Delta$\tablenotemark{f}}  & \colhead{$\alpha$\tablenotemark{g}} & \colhead{$\theta_{- \odot}$\tablenotemark{h}} & \colhead{$\theta_{-V}$\tablenotemark{i}} & \colhead{$\delta_{E}$\tablenotemark{j}}   }

\startdata

%2014 Oct 02 04:53 - 09:23 &  275 & D1.5  & 2.399 & 1.429 & 7.7   & 329.2 & 247.0 & 7.6 \\
2014 Feb 26 & 06:53 - 07:15 		& 57 & Keck & LRIS 	 &  6' x 6' & 0.135 & 11.875 & 11.463 & 4.4 & 79.0 & 261.2 & 0.15 \\
2016 Dec 09 & 12:22 - 12:48 		& 1074 &  KPNO &  HDI 	 & 	29' x 29' & 0.425 & 9.678 & 9.231 & 5.3 & 294.5 & 261.8 & 2.90 \\
2016 Dec 12 & 08:00 - 09:15 & 1077 		&  KPNO  & HDI 			 & 29' x 29' & 0.425 & 9.676 & 9.180 & 5.2 & 295.1 & 261.8 & 2.90 \\
2017 Mar 22 & 05:32 - 07:24 		& 1171 & KPNO &  HDI 		 & 29' x 29' & 0.425 & 9.605 & 8.877 & 4.2 & 95.8 & 261.2 & -1.04  \\
2017 Nov 15 & 	12:01 - 12:51		& 1415	& KPNO &  HDI 	 & 29' x 29'	& 0.425	& 	9.564 & 9.753 & 5.7 & 290.2 & 263.0 & 2.63 \\
2017 Nov 18 & 11:26 - 12:30		& 1418	& KPNO &  HDI 	 & 29' x 29'	& 0.425 &       9.564 & 9.708  & 5.8 & 290.6 & 263.0 & 2.69 \\
%2018 Mar 09 & 07:53  - 10:56 		& 1529 & WIYN &  ODI 	 & 60' x 60' & 0.11 & 9.609 & 8.642 & 1.4 & 98.9 & 263.1 & -0.37  \\
2018 Mar 11 & 23:23 - 02:10 		& 1531 & NOT &  ALFOSC 	 &  6' x 6' & 0.214 & 9.611 & 8.773 & 3.3 & 101.3 & 263.1 & -0.46 \\
2018 Apr 18 & 08:50 - 09:43 		& 1569 & Keck &  LRIS 	 &  6' x 6' & 0.135 & 9.635 & 9.022 & 4.9 & 109.7 & 263.0 & -2.19 \\

\enddata

%% Text for table notes should follow after the \enddata but before
%% the \end{deluxetable}. Make sure there is at least one \tablenotemark
%% in the table for each \tablenotetext.

\tablenotetext{a}{Telescope: Keck = Keck 10m, KPNO = KPNO 0.9m. %WIYN = WIYN 3.5m, 
NOT = NOT 2.5m}
\tablenotetext{b}{Instrument: LRIS = Low Resolution Imaging Spectrometer, HDI = Half-Degree Imager,  ALFOSC = Andalucia Faint Object Spectrograph and Camera}
\tablenotetext{c}{Field of view, in arcminutes}
\tablenotetext{d}{Image scale, arcseconds per pixel}
\tablenotetext{e}{Heliocentric distance, in AU}
\tablenotetext{f}{Geocentric distance, in AU}
\tablenotetext{g}{Phase angle, in degrees}
\tablenotetext{h}{Position angle of projected anti-solar direction, in degrees}
\tablenotetext{i}{Position angle of negative projected orbit vector, in degrees}
\tablenotetext{j}{Angle of Earth above orbital plane, in degrees}

\end{deluxetable}

\clearpage

\begin{deluxetable}{lccccccc}
%\tabletypesize{\scriptsize}
%\rotate

\tablecaption{Fixed-Aperture Photometry\tablenotemark{a} 
\label{photometry}}
\tablewidth{0pt}

\tablehead{ \colhead{UT Date} & DOY\tablenotemark{b} & $m_R(40)$\tablenotemark{c} & H$_{R}(40)$\tablenotemark{d} & $C_e(40)$\tablenotemark{e} &$\theta_N$\tablenotemark{f} & $\theta_S$\tablenotemark{f} & $\theta_{N-S}$\tablenotemark{g}}
\startdata
2014 Feb 26  & 57 		&   18.97$\pm$0.02	& 8.12 & 0.85 & 357$\pm 5$ & 175$\pm2$ 		& 182$\pm$5 \\
2016 Dec 09 & 1074 	&   17.06$\pm$0.10 	& 7.09 & 2.19 & 350$\pm2$ & 178$\pm1$ 	& 172$\pm$2\\
2016 Dec 12 & 1077 	&   16.91$\pm$0.04  &  6.96 & 2.47  & 352$\pm$1 & 174$\pm2$ 	& 178$\pm$2 \\
2017 Mar 22 & 1171 		&    16.68$\pm$0.01 	& 6.86 & 2.70 &   351$\pm$1 & 175$\pm$1 	& 176$\pm$2\\
2017 Nov 15 & 1415 		& 17.02$\pm$0.03    & 6.94 & 2.51   & 351$\pm$1 & 177$\pm$1 & 174$\pm$2\\
2017 Nov 18 & 1418  		& 17.06$\pm$0.03    & 6.95 & 2.49 & 350$\pm$1 & 178$\pm$1 & 172$\pm$2\\
2018 Mar 11  & 1531 	&   16.71$\pm$0.03	& 7.06  & 2.25 & 350$\pm$1 & 174$\pm$1 	& 176$\pm$2\\
%2018 Mar 09  & 1529	&  --- 	& --- &   --- & 351$\pm$1 & 173$\pm$1 & 178$\pm$1\\
2018 Apr 18  & 1569 	&   16.98$\pm$0.10	& 7.09  & 2.19 & 353$\pm$1 & 173$\pm1$ 	& 180$\pm$2\\

\enddata

%% Text for table notes should follow after the \enddata but before
%% the \end{deluxetable}. Make sure there is at least one \tablenotemark
%% in the table for each \tablenotetext.
\tablenotetext{a}{Projected aperture radius 4$\times$10$^4$ km at the comet}
\tablenotetext{b}{Day of Year, 1 = UT 2014 January 01}
\tablenotetext{c}{Apparent red magnitude within a 4$\times$10$^4$ km radius projected aperture}
\tablenotetext{d}{Absolute red magnitude computed from Equation (\ref{abs})}
\tablenotetext{e}{Scattering cross-section  in units of 10$^4$ km$^2$,  from Equation (\ref{area})}

\tablenotetext{f}{$\theta_N$ and $\theta_S$ are the measured position angles of the northern and southern arms of the coma, respectively, in degrees.}
\tablenotetext{g}{Absolute difference between $\theta_{N-S} = |\theta_N - \theta_S |$, in degrees.}

\end{deluxetable}

\clearpage

\clearpage

\clearpage

\clearpage

\clearpage

\clearpage

\clearpage

\clearpage

\clearpage

\clearpage

%%% The following command ends your manuscript. LaTeX will ignore any text

\end{document}